# High Thermoelectric Power Factor in Graphene/hBN Devices

Junxi Duan,[1,2,3] Xiaoming Wang,[2] Xinyuan Lai,[1] Guohong Li,[1] Kenji Watanabe,[4] Takashi Taniguchi,[4] Mona Zebarjadi,[2,3,*] Eva Y. Andrei[1,3,*]

[1]Department of Physics and Astronomy, Rutgers University, Piscataway, New Jersey 08854, USA.

[2]Department of Mechanical and Aerospace Engineering, Rutgers University, Piscataway, New Jersey 08854, USA.

[3]Institute of Advanced Materials, Devices, and Nanotechnology, Rutgers University, Piscataway, NJ, 08854, USA.

[4]Advanced Materials Laboratory, National Institute for Materials Science, 1-1 Namiki, Tsukuba 305-0044, Japan.

**ABSTRACT**

Fast and controllable cooling at nanoscales requires a combination of highly efficient passive cooling and active cooling. While passive cooling in graphene-based devices is quite effective due to graphene's extraordinary heat-conduction, active cooling has not been considered feasible due to graphene's low thermoelectric power factor. Here we show that the thermoelectric performance of graphene can be significantly improved by using hBN substrates instead of $SiO_2$. We find the room temperature efficiency of active cooling, as gauged by the power factor times temperature, reaches values as high as 10.35 $Wm^{-1}K^{-1}$, corresponding to more than doubling the highest reported room temperature bulk power factors, 5 $Wm^{-1}K^{-1}$ in $YbAl_3$, and quadrupling the best 2D power factor, 2.5 $Wm^{-1}K^{-1}$, in $MoS_2$. We further show that in these devices the electron-hole puddles region is significantly reduced. This enables fast gate-controlled switching of the Seebeck coefficient polarity for applications in n- and p-type integrated active cooling devices.

As the size of the electronic components shrinks, larger power densities are generated, resulting in local hot spots. The small size of these hot spots and their inaccessibility make it difficult to maintain a low and safe operating temperature.[1] Solid-state integrated active thermoelectric coolers could solve the long lasting electronic cooling problem.[2, 3] In the normal refrigeration mode of thermoelectric coolers, heat is pumped from the cold side to the hot side. However, there is an increasing need to pump heat from the hot spots generated on the chip to the colder ambient reservoir. In this mode of operation, both passive and active cooling can be used.[4] In the case of passive cooling, where heat is transported via the phonon channel, the performance is fixed by the thermal conductance. In contrast, active cooling which uses the Peltier effect to pump heat via the electronic channel can be controlled and tuned with applied current. The performance of Peltier cooling is a function of the thermoelectric power factor, $PF = \sigma S^2$, where $\sigma$ is the electrical conductivity and $S$ is the Seebeck coefficient. In this manuscript, we also use the notation of $PFT$, referring to $PF$ times temperature $T$ which has a more convenient unit of Wm$^{-1}$K$^{-1}$ (same as thermal conductivity). Although there is no theoretical limit on $PF$, the interplay between the Seebeck coefficient and the electrical conductivity in highly doped bulk semiconductors, has so far prevented the realization of very large thermoelectric power factors.[5-7]

Single-layer graphene possesses extraordinary electronic and thermal properties.[8-12] In particular its higher mobility, which due to the weak electron-phonon interaction persists up to room temperature, can be orders of magnitude higher than in other 2D thermoelectric materials, such as semiconducting transition metal dichalcogenides (TMDs).[13-16] Theoretical and experimental studies show that the Seebeck coefficient in graphene could reach values comparable to that in bulk semiconductors by decreasing the carrier density.[17-23] Both the Seebeck coefficient and the mobility play an important role in active cooling. At the same time graphene's extremely large thermal conductivity also enables efficient passive cooling.[12] Furthermore, the ability to control its carrier density by electrostatic gating rather than by chemical doping imparts graphene an important advantage over bulk materials.

As a purely 2D material, the electronic properties of graphene are severely affected by its surroundings. Experiments demonstrate that the most commonly used $SiO_2$ substrate has many surface charged states and impurities which cause strong Coulomb scattering that limits the mobility and introduces large potential fluctuations in G/$SiO_2$ samples.[24-26] The potential fluctuations induce electron-hole puddles (EHP) in the vicinity of the charge-neutrality point (CNP) and prevent gating for lower carrier density.[25] Depositing graphene on hBN substrates, which are relatively inert and free of surface charge traps, produces G/hBN samples with smaller potential fluctuations and higher mobility than G/$SiO_2$.[27-29] Here we report on measurements of the thermoelectric properties, $S$ and $PF$, for G/hBN and G/$SiO_2$ samples.

Fig. 1a shows a schematic of the apparatus, which allows measuring both electrical and thermal transport properties of the material (see Supporting Information). Fig 1b shows the Seebeck coefficient measured in G/hBN and G/$SiO_2$ samples. In G/hBN sample, the peak value $S = 182\mu V/K$ at 290K is significantly higher than the corresponding value $S = 109\mu V/K$ in the G/$SiO_2$ sample, and the gate voltage at the peak position, $V_p = -2.2V$, is closer to the CNP than that in the G/$SiO_2$ sample, $V_p = -4.5V$. From the measured values of $S$ and the conductivity we calculate the value of $PFT=S^2\sigma T$ as a function of carrier density shown in Fig. 1c.[30] The *PFT* first increases with decreasing carrier density when far from CNP, then after reaching a peak value, it drops to zero at the CNP. We find that the room temperature peak value of *PFT* in G/hBN, 10.35 $Wm^{-1}K^{-1}$, is almost twice that in G/$SiO_2$, 6.16 $Wm^{-1}K^{-1}$. This value is larger than the record value in bulk materials at room temperature reported for $YbAl_3$ (~5 $Wm^{-1}K^{-1}$), and larger than the value at room temperature in 2D materials reported for $MoS_2$ (~2.5$Wm^{-1}K^{-1}$) and $WSe_2$ (~1.2$Wm^{-1}K^{-1}$).[30-33] We note that this value of the *PFT* is in fact underestimated since, due to the two-probe measurement of the conductivity, the contact resistance is included in the conductivity calculation. As we discuss later the *PFT* value increases with temperature and has not yet saturated at room temperature. Therefore, even larger *PFT* values are expected at higher temperatures.

We next use the linear Boltzmann equation in the relaxation time approximation to relate the Seebeck coefficient to the experimentally controlled quantities. Within this model the response of the electrical and thermal current densities, $j$ and $j_q$, to the electric field, $E$, and temperature gradient, $\nabla T$, are given by:[17]

$$j = L^{11}E + L^{12}(-\nabla T) \qquad (1)$$

$$j_q = L^{21}E + L^{22}(-\nabla T) \qquad (2)$$

where $L^{11} = K^{(0)}, L^{12} = -\frac{1}{eT}K^{(1)}, L^{21} = -\frac{1}{e}K^{(1)}, L^{22} = \frac{K^{(2)}}{e^2T}$, and

$$K^{(m)} = \int_{-\infty}^{+\infty} d\epsilon (\epsilon - \mu)^m \left(-\frac{\partial f^0(\epsilon)}{\partial \epsilon}\right)\sigma(\epsilon) \quad m = 0,1,2 \qquad (3).$$

Here, $\epsilon(k) = \hbar v_F k$, $v_F$ is the Fermi velocity, $\mu$ is the chemical potential, $f^0(\epsilon)$ is the equilibrium Fermi-Dirac distribution function. The differential conductivity is $\sigma(\epsilon) = e^2 v_F^2 \frac{D(\epsilon)\tau(\epsilon)}{2}$, $D(\epsilon) = 2|\epsilon|/(\pi\hbar^2 v_F^2)$ is the density of states including the 4-fold degeneracy of graphene, $v_F = 10^6 \text{ms}^{-1}$ is the Fermi velocity, and $\tau(\epsilon)$ is the relaxation time.[34] The Seebeck coefficient is defined as $S = L^{12}/L^{11}$, the electrical and thermal conductivity are $\sigma = L^{11}$ and $\kappa = L^{22}$ respectively, and the Peltier coefficient is $\Pi = L^{21}/L^{11}$.[17, 35] Importantly, we note that the Seebeck coefficient is controlled by the energy dependence of the conductivity.

In Fig. 1d we show the calculated carrier density dependence of the Seebeck coefficient at 300K in the presence of random potential fluctuations induced by a distribution of charge impurities (see Supporting Information). The calculation follows the model proposed in Ref. 15 and, for simplicity, considers only the screened Coulomb scattering which is known to be the dominant scattering mechanism in this system.[15, 17, 36-38] We note that the monotonic increase of $S$ with decreasing carrier density peaks at the point where the Fermi energy enters the EHP region.[17, 18] In this region (shadow area in Fig. 1d) both electrons and holes are present, but since they contribute oppositely to $S$, the value of $S$ drops. Consequently, the smaller the EHP region, the higher the peak value of $S$. There is, however, a limit to the magnitude of $S$ that is set by the temperature. When $k_B T$ is comparable to the potential fluctuations energy scale, the peak value of $S$ is controlled by the temperature. The effect of inserting the hBN spacer, typically $d\sim10$nm, is to increase the distance from the charge impurities in the SiO$_2$ substrate which reduces the magnitude of the random potential fluctuations in the graphene plane. This reduces the EHP region and, as a consequence, results in a larger value of $S$ (see Supporting Information). Again, there is a limit to this improvement. In the limit of infinitely large separation, *i.e.* no Coulomb scattering, thermally excited phonons become the dominant mechanism which limits the value of $S$. In the acoustic phonon-dominated regime, the Seebeck coefficient at room temperature is expected to be smaller than $S = 100\mu V/K$.[17]

As discussed above, the peak position of $S$ marks the boundary of the EHP region, which depends on both the temperature and the extent of the random potential fluctuations. In the high temperature limit this region is dominated by thermal excitations, while at low temperatures it is controlled by the energy scale of the random potential fluctuations. Currently, most measurements of the EHP are carried out by scanning probe microscopy, which are typically performed at low temperatures and over a scanning range much smaller than normal transport devices.[28, 29, 39] Although the size of the EHP region can be estimated from the gate dependence of the resistivity,[27] the peak position of $S$ provides a more direct measure of the EHP region. In Fig. 2a, showing the back-gate dependence of $S$ in the temperature range from 77K to 290K, we note that as temperature decreases so does the peak value of $S$ and its position, $V_P$, moves closer to the CNP. In the following discussion, we focus on the hole side since the peaks on this side are clearer in the G/SiO$_2$ sample. The temperature dependence of $V_P$, shown in Fig. 2b for both samples, follows an exponential function $V_P(T) = a + b(e^{\alpha T} - 1)$ where $a$, $b$ and $\alpha$ are fitting parameters. The intercept $a$ at $T=0$ is 0.12V and 0.52V corresponding to density fluctuations of $1.8 \times 10^{10}$cm$^{-2}$ and $7.6 \times 10^{10}$cm$^{-2}$ for the G/hBN and G/SiO$_2$ samples respectively. Both values are comparable to previous results measured by scanning tunneling microscopy at liquid-helium temperature.[29] The corresponding energy scale of the random potential fluctuations in the two samples is 21.8meV and 45.4meV, respectively. Seebeck coefficient peak positions extracted from previous studies are also shown.

Unlike the case of the voltage drop in electrical transport, which is insensitive to the sign of the carrier charge, the Seebeck voltage reverses its sign when switching from hole-doping to electron-doping. In the G/hBN sample the polarity of the peak Seebeck coefficient could be reversed with a relatively small gate voltage ~$2V_P$. We define the slope of this polarity-switching effect as $\beta = S_p/V_p$, where $S_p$ stands for the peak Seebeck coefficient. In Fig. 2c, $\beta$ in G/hBN and G/SiO$_2$ samples at different temperatures are shown together with values extracted from previous studies in G/SiO$_2$ samples. Clearly, the value of $\beta$ is strongly enhanced in G/hBN sample.

The ambipolar nature of graphene, which allows smooth gating between electron and hole doped sectors, together with the large values of $\beta$ which facilitate switching the polarity of *S*, extend a distinct advantage in applications where p-type and n-type devices are integrated. This can be seen in the thermoelectric active cooler design shown in Fig. 2d, which can pump heat from the hot end ($T_H$) to the cold end ($T_L$) in a controlled and fast manner using combined active and passive cooling. In this G/hBN based device, the p-n legs are arranged thermally in parallel and electrically in series to maximize the active cooling.[4] Its structure, which is readily realized with lithographically patterned gates is significantly simpler than that of bulk devices that require different materials or different doping for the p and n legs. At the optimal value of applied current, the active cooling power is $P_{active} = PFT_H \cdot T_H/2$.[4] On the other hand, the passive cooling power is $P_{passive} = \kappa \Delta T$ where $\kappa \sim 600 \text{ Wm}^{-1}\text{K}^{-1}$ is the thermal conductivity of graphene supported on a substrate at room temperature.[12] For $T_H = 330K$ and $\Delta T = 30K$, active cooling contributes an additional 10% over the passive cooling. At higher temperatures, as *PFT* increases and thermal conductivity decreases, the contribution of active cooling increases further.

The temperature dependence of the Seebeck coefficient at a fixed back gate voltage for both samples is shown in Fig. 3a. The corresponding carrier density in G/hBN and G/SiO$_2$ is $2.0 \times 10^{12} \text{cm}^{-2}$ and $3.0 \times 10^{12} \text{cm}^{-2}$, respectively. Peak Seebeck coefficient values extracted from previous studies are also presented and show similar values as in our G/SiO$_2$ sample. The Seebeck coefficient values measured in both devices show nonlinear temperature dependence. Indeed in the case of screened Coulomb scattering, the temperature dependence of the Seebeck coefficient is quadratic rather than linear.[15] Using this model (see Supporting Information), we calculate the temperature dependence of the Seebeck coefficient from the general Mott's formula, as shown in Fig. 3a. The temperature dependence of the measured and calculated *PFT* is also shown in Fig. 3b together with a comparison with values extracted from previous studies. The calculation overlaps with experimental results quite well. In hBN encapsulated graphene samples with much higher mobility, the violation of the general Mott's formula in graphene was recently attributed to inelastic electron-optical phonon scattering.[40] For the devices reported here, the good agreement between the experiment and calculation suggests that general Mott's formula is still valid and screened Coulomb scattering is dominant.

In summary, the conductivity and Seebeck coefficient are measured in G/hBN and G/SiO$_2$ samples in the temperature range from 77K to 290K. At room temperature, the peak Seebeck coefficient in G/hBN reaches twice the value measured in G/SiO$_2$ and the peak *PFT* value reaches 10.35 Wm$^{-1}$K$^{-1}$, which significantly exceeds previously reported records in both 2D and 3D thermoelectric materials. In G/hBN we find that the density fluctuations due to the substrate induced random potential fluctuations, $1.8 \times 10^{10}$cm$^{-2}$, represents a four-fold reduction compared to the value in G/SiO$_2$ sample $7.6 \times 10^{10}$cm$^{-2}$. Our findings show that the fast and low-power bipolar switching make it possible to integrate all-in-one graphene p-type and n-type devices. The study demonstrates the potential of graphene in thermoelectric applications especially in electronic cooling where large thermal conductivity (passive cooling) and large thermoelectric power factor (active cooling) are needed simultaneously.

## ASSOCIATED CONTENT

**Supporting Information**

Device fabrication, Seebeck measurement, summary of recently reported *PFT* in 2D materials, nonlinear dependence on temperature, results from other G/hBN samples, scattering mechanism, and calculation details.

## AUTHOR INFORMATION

**Corresponding Author**

*Correspondence and requests for materials should be addressed to M.Z. (email: m.zebarjadi@rutgers.edu) and E.Y.A. (email:eandrei@physics.rutgers.edu).

**Author contributions**

The manuscript was written through contributions of all authors. All authors have given approval to the final version of the manuscript.

**Notes**

The authors declare no competing financial interest.

## ACKNOWLEDGEMENTS

M. Z. and J. X. D. would like to acknowledge the support by the Air Force young investigator award, grant number FA9550-14-1-0316. E. Y. A. and J. X. D. would like to acknowledge support from DOE-FG02-99ER45742 and NSF DMR 1207108.

**Figure captions**

**Figure 1. Thermoelectric measurement of Graphene at room temperature.** (a) Optical micrograph of the graphene on hBN (G/hBN) device. (b) Measured Seebeck coefficient in G/hBN and G/SiO$_2$ devices as a function of back gate at 290K. Inset: measured resistance in both devices at 290K. (c) Measured *PFT* in both samples as a function of back gate at 290K. (d) Simulation of carrier density dependence of the Seebeck coefficient at 300K using the screened Coulomb scattering model for two values of the hBN thickness, *d*, and random potential fluctuations, $E_{RP}$, induced by charge impurities (See Supporting Information). The rectangular shadow corresponds to the EHP region in a sample with *d*=10nm, and $E_{RP}$=40meV.

**Figure 2. Temperature dependence of Seebeck coefficient and EHP region.** (a) Measured Seebeck coefficient in the G/hBN device as a function of back gate and temperature. (b) Temperature dependence of peak positions of the Seebeck coefficient ($V_p$) on the hole side for G/hBN (solid squares) and G/SiO$_2$ (open squares) devices are shown together with the exponential fit discussed in the text (solid lines). (c) Slope of polarity-switching effect from both our devices (solid squares for G/hBN and open squares for G/SiO$_2$). Values of $V_p$ and slope in G/SiO$_2$ samples (open triangles) extracted from previous studies are also shown. (d) Sketch of the active cooler with integrated n-type and p-type legs.

**Figure 3. Temperature dependence of Seebeck coefficient and *PFT* at fixed carrier density.** (a) Measured Seebeck coefficient in G/hBN (solid squares) and G/SiO$_2$ (open squares) devices are plotted together with the theoretical values (solid lines) calculated by using the screened Coulomb scattering model discussed in the text. Dashed lines serve as guides to emphasize the nonlinear behavior. (b) Measured *PFT* (solid and open squares) from both devices are compared with theoretical values (solid lines). Peak Seebeck coefficient and *PFT* values extracted from previous studies in G/SiO$_2$ samples (open triangles) are also presented.

**Figure 1**

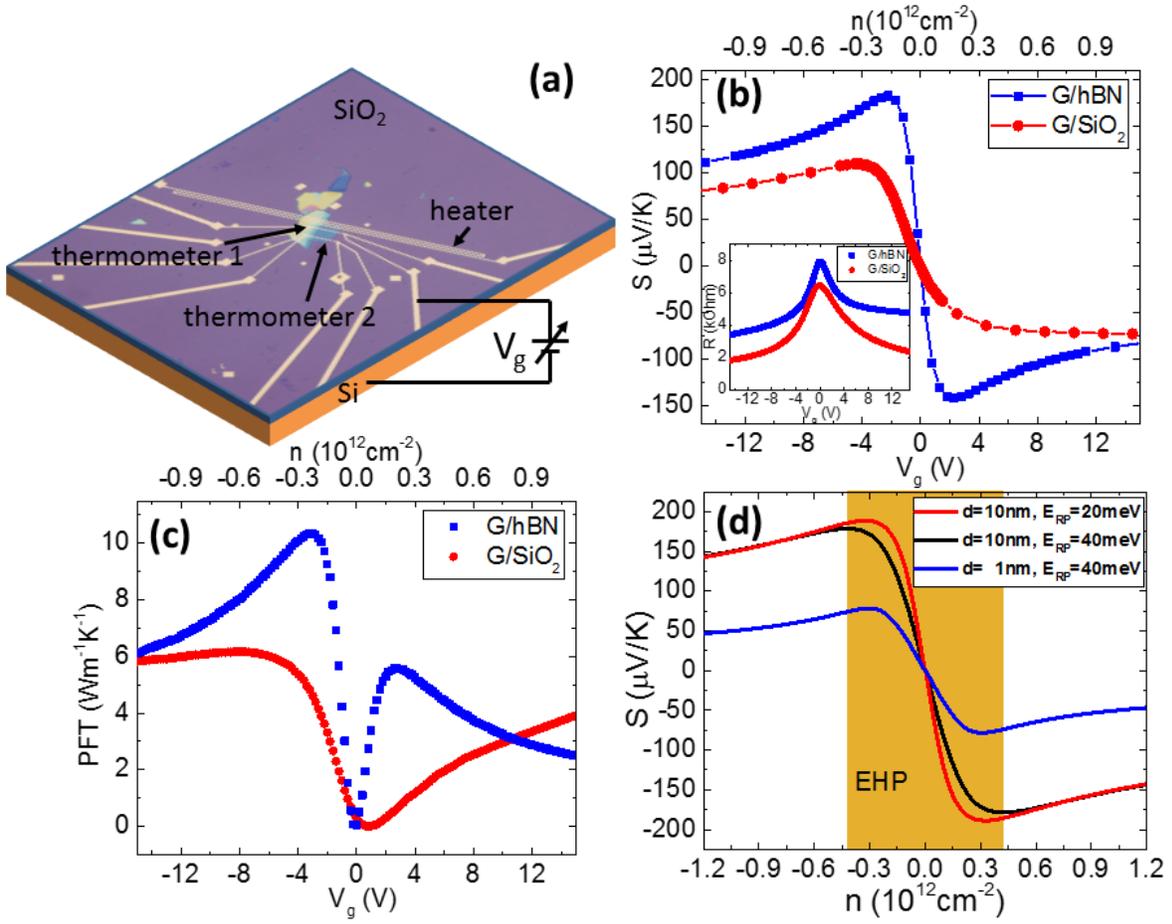

**Figure 2**

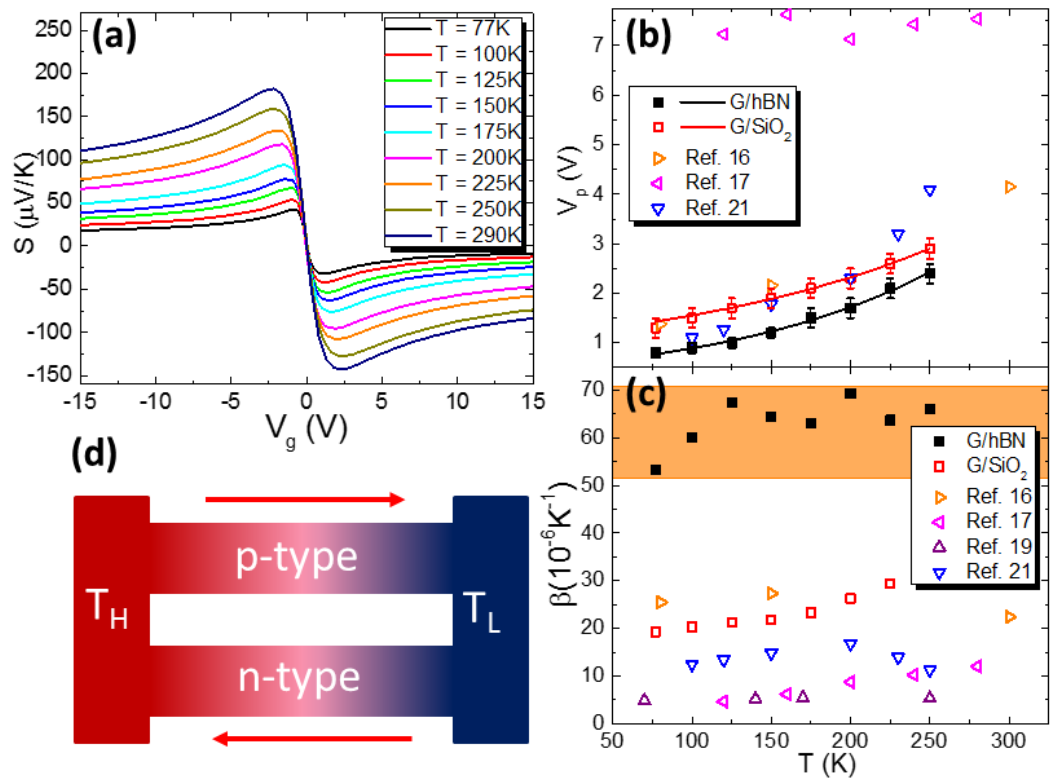

**Figure 3**

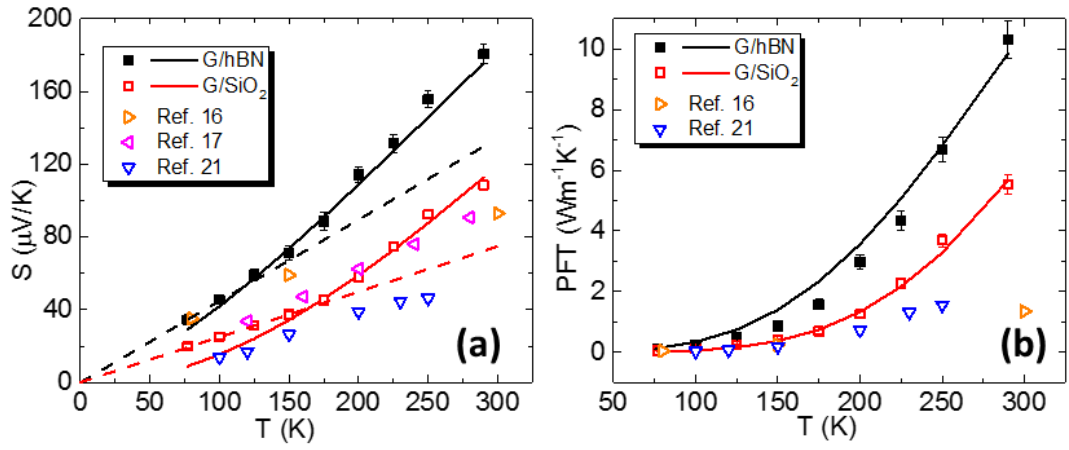

## SUPPORTING INFORMATION

**Device fabrication**

Graphene on hBN samples are fabricated using the PMMA-based dry transfer method. hBN is exfoliated on 300nm SiO$_2$/Si substrate. Thickness of hBN is measured by atomic-force microscope (AFM). The number of the device shown in the test is 10nm. Single-layer graphene is prepared on PMMA membrane. It could be distinguished through the color difference under optical microscope and later identified by AFM after transfer. Graphene on SiO$_2$ samples is directly exfoliated on SiO$_2$ surface. Electrodes on graphene serve as voltage probes and thermometers measuring the local temperature at the two ends of graphene flake. A strip of gold wires beside the sample is used as heater. Figure 1a shows the optical image of a typical sample. To induce a uniform temperature gradient across the sample, the size of the heater (400 μm) is much larger than the size of graphene flake (typically 20μm × 10μm) as well as thermometers (40 μm) on it. Both the thermometers and the heater are defined by standard electron beam lithography. Cr/Au (3/45 nm) layers are deposited on it using electron beam evaporation. All the samples are annealed in forming gas (H$_2$/Ar) at 230°C over 12 hours to remove resist residue before measurements. We measured three G/hBN (S1-S3, S1 is the one shown in the text) and one G/SiO$_2$ samples in total.

In the main text, we show the results from G/hBN S1 and G/SiO$_2$ samples with the highest mobility, $1.8 \times 10^4 \text{cm}^2\text{V}^{-1}\text{s}^{-1}$ and $1.2 \times 10^4 \text{cm}^2\text{V}^{-1}\text{s}^{-1}$ on the hole side at 77K respectively. To calculate the carrier density, we adopt the parallel-plates-capacitor model. The thickness of used SiO$_2$ is 300nm and thickness of used hBN is determined by AFM. The dielectric constant of SiO$_2$ and hBN are known. By measuring the resistance at different back gate voltages, one can determine the position of the charge-neutrality point which corresponds to the resistance peak. One can then take it as the n=0 point and obtain the charge carrier concentration with the calculated capacitance and the voltage difference from the charge-neutrality point.

**Seebeck measurement**

Temperature is measured through 4-probes resistance measurement of the thermometers with resolution smaller than 0.01K. By powering up the heater, a temperature gradient $\Delta T$ is generated along the sample (See figure 1a). The thermally-induced voltage $\Delta V$, is measured by the voltage probes at the two ends of the sample. The Seebeck coefficient is then calculated using: $S = \Delta V/\Delta T$. $\Delta T \ll T$ is required to make the measurement in the linear response region. All the measurements are done in vacuum ($P \sim 10^{-6}$Pa) with a temperature range from 77K to 300K.

To minimize the systematic error caused by the non-uniform temperature distribution along the thermometer lines, the length of the heater (400μm) is designed to be an order of magnitude longer than the thermometers (40μm) and the graphene channel (typically 20μm). Figure S1a and S1b show the simulated temperature distribution in the device. The heater locates in the middle of the device. Since the thermal conductivity of the Au heater is two orders of magnitude higher than $SiO_2$, a uniform heater temperature (T=305K) was assumed. The temperature far away from the heater (T=300K) was taken as the reference point. Thermal conductivity of $SiO_2$ at room temperature (about 2W/mK) was used. Figure S1c shows isotherms close to the heater. Since the size of the thermometers and channel is only 1/10 of the size of heater, one can see that the isotherms in this small region (as indicated) are actually straight lines running parallel to the heater. From the simulation, one can estimate that the systematic error introduced by the finite thermometer size is less than 0.1%.

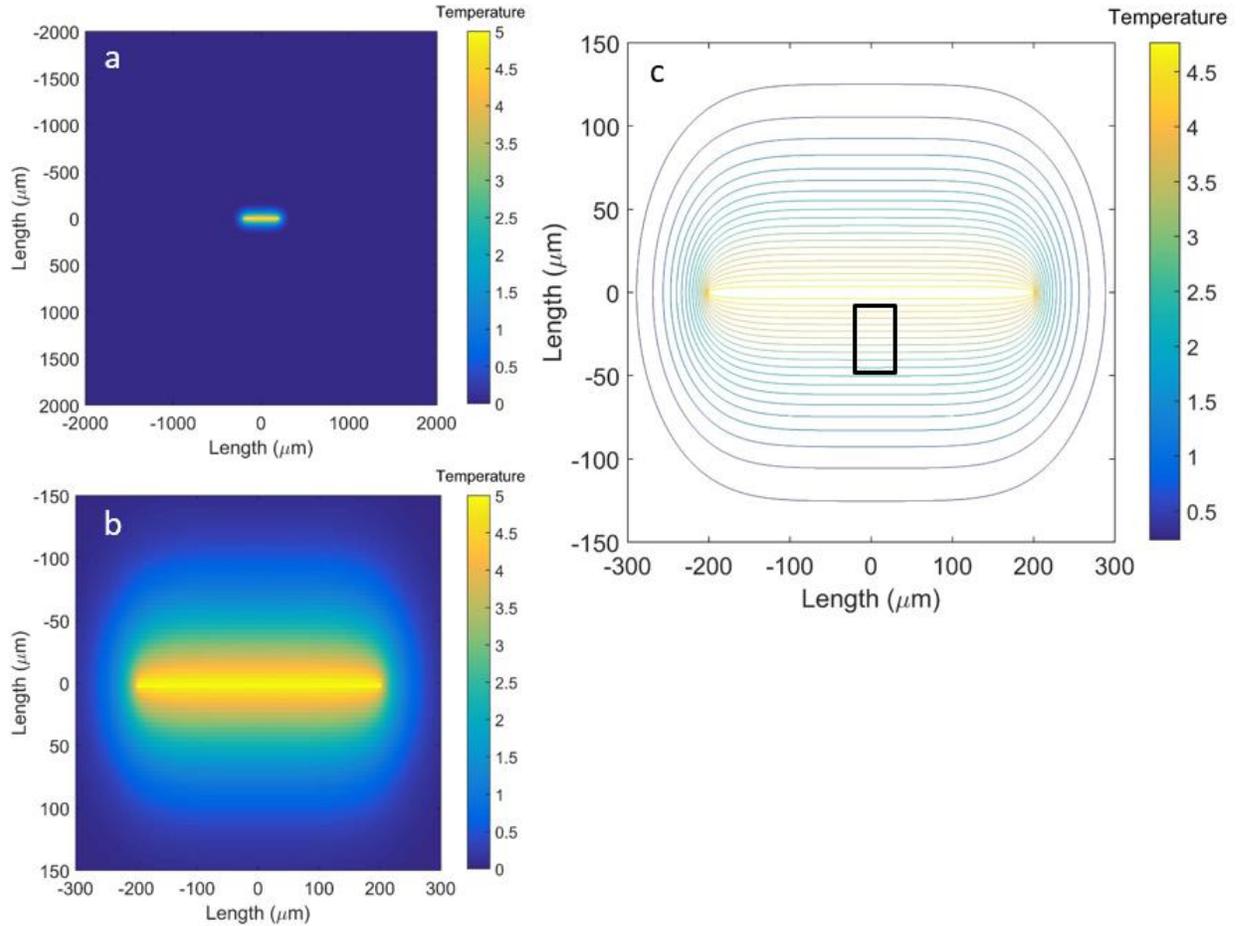

**Figure S1. Simulation of temperature distribution.** (a) Temperature distribution in the device (4000μm × 4000μm). (b) zoom-in view of panel a. The temperature of the 400μm heater in the center is assumed to be uniform. (c) Isotherms in the vicinity of the hater. The rectangular box 40μm × 40μm indicates the area where the thermometers and graphene channel are located in.

**Summary of recently reported *PFT* in 2D materials**

Comparing to other 2D materials, the G/hBN samples in this study show much higher *PFT* at room temperature. Table S1 summarizes recently reported optimized thermoelectric power factor times temperature (*PFT*) in 2D materials at room temperatures. Also, to make the comparison clear, Figure S2 shows the carrier density dependence of conductivity, Seebeck coefficient and *PFT* of the G/hBN sample.

| Sample | G/hBN | G/SiO$_2$ | WSe$_2$ (3L) | MoS2 (2L) | MoS2 (1L) |
|---|---|---|---|---|---|
| **PFT**(Wm$^{-1}$K$^{-1}$) | 10.35 | 6.16 | 1.2 | 2.5 | 0.9 |
| **Ref.** | this study | this study | 32 | 30 | 31 |

**Table S1. A summary of recently reported *PFT* at room temperature in 2D materials.**

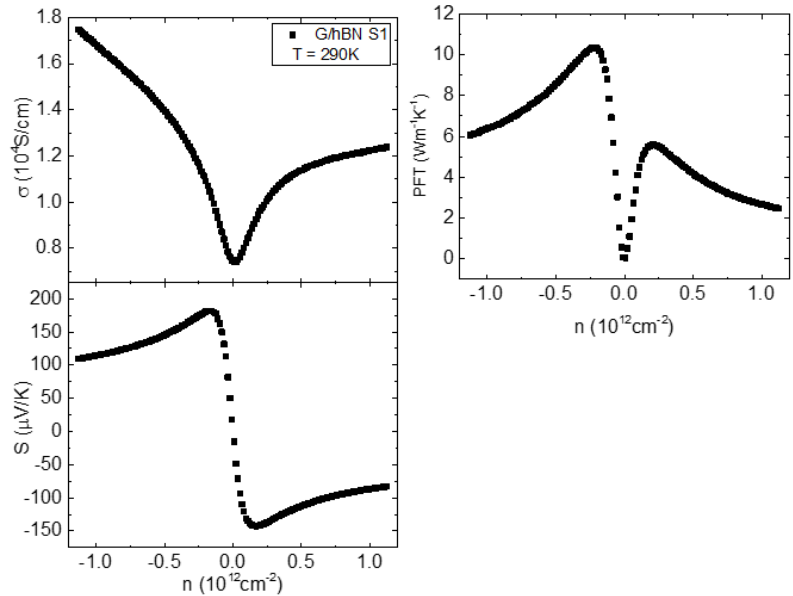

**Figure S2. Carrier density dependence of conductivity, Seebeck coefficient and *PFT* in G/hBN S1.**

**Nonlinear temperature dependence**

The nonlinearity of the measured Seebeck coefficient versus temperature of both G/hBN and G/SiO$_2$ samples with both low and high carrier density is shown in Figure S4. Dashed lines are guides indicating the nonlinearity. All cases, high and low density in G/hBN and G/SiO$_2$ samples, show deviations from linear temperature dependence above certain temperature (about 100K).

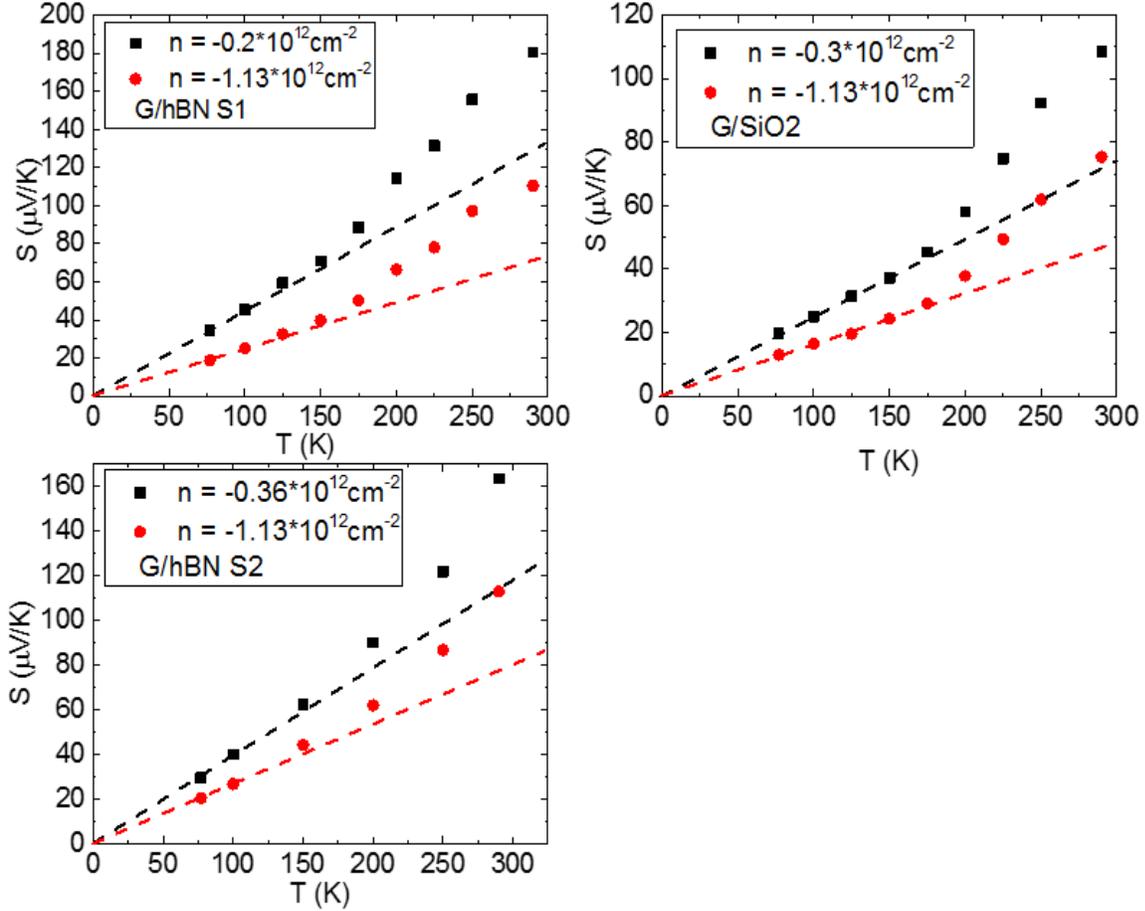

**Figure S3. Seebeck coefficient in three samples with different carrier densities versus temperature.**

**Results from other G/hBN samples**

We have measured other G/hBN samples whose results are not shown in the main text. The Seebeck coefficient measured at room temperature in these samples is shown below in Figure S5.

Samples G/hBN S1 and G/SiO$_2$ are shown in the main text. In G/hBN sample S2 and S3, the size of electron-hole puddle region as indicated by the position of $V_p$ is comparable to the G/SiO$_2$ sample. The Seebeck coefficient measured is larger than the result from G/SiO$_2$ sample. This suggests that by increasing the distance to the charge impurities the hBN spacer helps increase the Seebeck coefficient. By comparing the results among the three G/hBN samples we note that the increase in the peak value of the Seebeck coefficient is directly correlated with the decrease in $V_p$ This result demonstrates that by reducing the size of electron-hole puddles one can improve the Seebeck coefficient further. The thickness of hBN in the three G/hBN samples is in the range from 10nm to 20nm. The simulation shows that the Seebeck coefficient in G/hBN starts saturating when hBN spacer is thicker than 10nm.

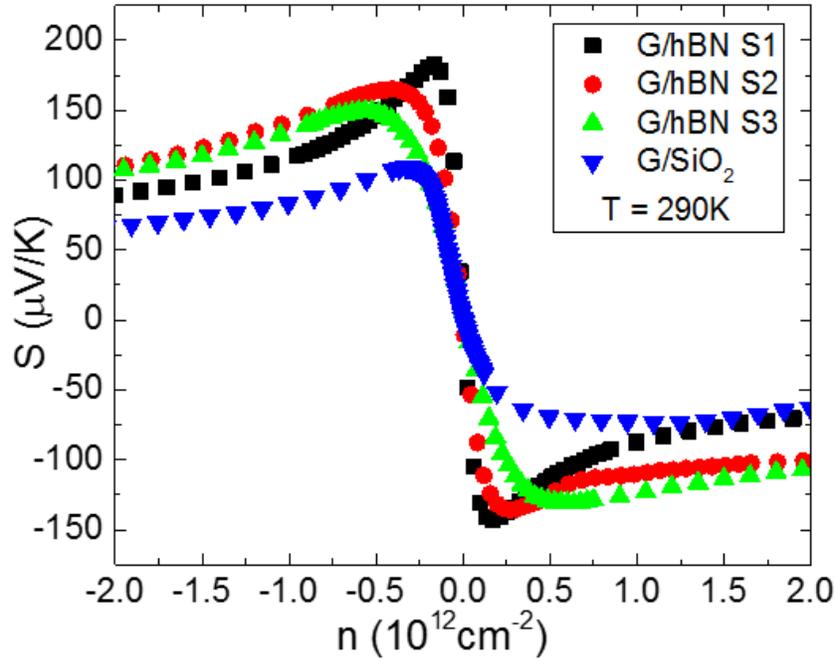

**Figure S4. Data from other G/hBN samples as well as the samples shown in the text.**

**Scattering mechanism**

In G/SiO$_2$ samples, previous studies show that the charge impurity scattering is the dominant mechanism. In the G/hBN samples studied, as shown in Figure S2, the conductivity dependence on back gate voltage show deviation from the linear behavior which indicate a crossover of the dominant scattering from Coulomb scattering at low carrier density to short-range impurity scattering at high carrier density.[27, 41, 42] In Figure 3, the density used in both samples is smaller than the crossover point which means that the dominant scattering mechanism is still charge impurity scattering, as supported by the good agreement between the experiment and calculation. Although hBN is used to reduce the density of charge impurities, imperfectness of hBN, contamination during the fabrication could introduce charge impurities and affect the performance of the device.

**Calculation details**

For screened Coulomb scattering, the energy dependence of the relaxation time is

$$\frac{1}{\tau(\epsilon_k)} = \frac{\pi n_i}{\hbar} \int \frac{d^2 k'}{(2\pi)^2} \left|\frac{v_i(q)}{\varepsilon(q,T)}\right|^2 \delta(\epsilon_k - \epsilon_{k'})(1 - cos\theta)(1 + cos\theta) \qquad (S1)$$

where $\theta$ is the scattering angle, and $v_i(q) = 2\pi e^2 \exp(-qd)/(\kappa q)$ is the Fourier transform of a 2D Coulomb potential with dielectric constant $\kappa$. The 2D finite-temperature static random phase approximation screening function is $\varepsilon(q,T) = 1 + v_c(q)\Gamma(q,T)$, where $\Gamma(q,T) = D_0 \tilde{\Gamma}(q,T)$ is the irreducible finite-temperature polarizability function, $D_0 = 2E_F/\pi\hbar^2 v_F^2$ is the DOS at Fermi level, and $v_c(q)$ is the bare Coulomb interaction. For $T \ll T_F$,

$$\tilde{\Gamma}(q,T) \approx \begin{cases} \frac{\mu}{E_F}, & \epsilon(q) < 2\mu, \\ \frac{\mu}{E_F}\left[1 - \frac{1}{2}\sqrt{1 - \left(\frac{2\mu}{\epsilon(q)}\right)^2} - \frac{\epsilon(q)}{2\mu}\sin^{-1}\frac{2\mu}{\epsilon(q)}\right] + \frac{\pi q}{8k_F} + \frac{2\pi^2}{3}\frac{T^2}{T_F^2}\frac{E_F\mu}{\epsilon^2(q)}\frac{1}{\sqrt{1 - (2\mu/\epsilon(q))^2}}, & \epsilon(q) > 2\mu, \\ \frac{\mu}{E_F} + \sqrt{\frac{\pi\mu}{2E_F}}\left(1 - \frac{\sqrt{2}}{2}\right)\zeta\left(\frac{3}{2}\right)\left(\frac{T}{T_F}\right)^{\frac{3}{2}}, & \epsilon(q) = 2\mu, \end{cases}$$

where $\zeta(x)$ is Riemann's zeta function. The Fermi velocity in graphene is taken as $1 \times 10^6$ m/s. $\kappa$ is taken as the average of the dielectric constants of the vacuum and the substrate which is (1+4)/2=2.5. The fine-structure parameter $r_s$ is 0.85.

In the linear response approximation, the Seebeck coefficient is defined as:

$$S = \frac{-1}{eT} \frac{\int_{-\infty}^{+\infty} d\epsilon (\epsilon - \mu)\left(-\frac{\partial f^0(\epsilon)}{\partial \epsilon}\right)\sigma(\epsilon)}{\int_{-\infty}^{+\infty} d\epsilon \left(-\frac{\partial f^0(\epsilon)}{\partial \epsilon}\right)\sigma(\epsilon)} \tag{S2}$$

$$\sigma(\epsilon) = e^2 v_F^2 D(\epsilon)\tau(\epsilon)/2 \tag{S3}$$

The Seebeck coefficient is not sensitive to the magnitude of the conductivity, but it is sensitive to its energy dependence. In the calculation of $S$, the integrand is done within a Fermi window of $[-10k_BT, +10k_BT]$ centered at Fermi energy $\mu$.

Charge impurities generate random potential fluctuations (RP) and scatter the carriers. Electron-hole puddles are the result from both temperature excitation and random potential fluctuations. To simulate the behavior with RP, a simplified model is adopted. The existence of RP will introduce carrier density fluctuations in the graphene. The whole graphene channel could be divided into small islands with uniform carrier density within each island. Total Seebeck coefficient will be the average of the contribution from all these islands weighted by the thermal conductivity and volume fraction of each island. Since the thermal conductivity of graphene is dominated by phonon, the thermal conductivity of each island could be taken the same.[43, 44] The total Seebeck coefficient will be the average of the Seebeck coefficient from all islands. The distribution of the potentials in these small islands could be taken as uniformly distributed in a range $[-E_{RP}, E_{RP}]$. In Figure 1d, the Seebeck coefficient calculated with larger $E_{RP}$ shows a lower but wider peak.

When inserting the hBN spacer between graphene and the $SiO_2$ substrate, the Coulomb potential in the graphene plane is weakened. If the density of charge impurities is unchanged, the potential fluctuations will become smaller. The simulation shows stronger electron-hole asymmetry in the contribution (Z) to the Seebeck coefficient in each island with uniform carrier density, as shown in Figure S6 with $n = 2 \times 10^{15} \text{m}^{-2}$, where $Z = (\epsilon - \mu)\left(-\frac{\partial f^0(\epsilon)}{\partial \epsilon}\right)\sigma(\epsilon)/\Lambda$ and $\Lambda$ is a normalization parameter. This asymmetry increases the Seebeck coefficient.

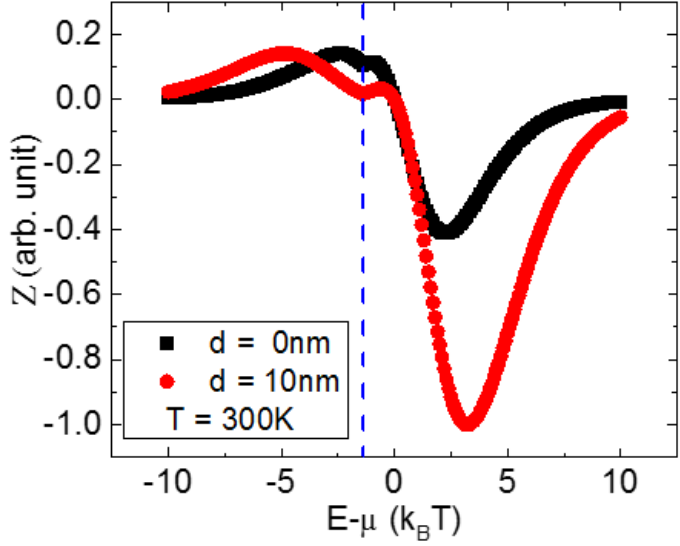

**Figure S5. Comparison of electron-hole asymmetric contribution (Z) to Seebeck coefficient between *d*=0nm and *d*=10nm calculations with carrier density of $2.0 \times 10^{15}$m$^{-2}$. Dash line gives the position of CNP.**

Figure 1d shows that the distance *d* between the graphene layer and the charge impurities is a crucial parameter. To simulate the temperature dependence, we choose the carrier density to be outside the EHP region where the effect of RP could be neglected, as shown in Figure 1d. For the G/SiO$_2$ sample, the charge impurities are on the SiO$_2$ surface. Assuming the same distance, *d*, for all the impurities we find that for with $d = 1.1$nm and $n_i = 3.2 \times 10^{15}$m$^{-2}$ the simulation agrees with the experiment. However, for the G/hBN device, if we calculate the Seebeck coefficient in the same manner, the temperature dependence is quite different from what we measured. The discrepancy can be attributed to the fact that in the G/hBN sample, there are also charge impurities located on the graphene/hBN interface. The simulation for the G/hBN device therefore uses two layers of charge impurities. As shown in Fig. 3 of the main text, that using parameters $d_1 = 10$nm, $n_{i1} = 5 \times 10^{15}$m$^{-2}$ and $d_2 = 0$nm, and $n_{i2} = 1 \times 10^{15}$m$^{-2}$ for the layer distance and impurity density in layer 1 and layer 2 respectively matches the experimental results.

The small deviations of the experimental data from the theoretical prediction can be attributed to several simplifications. First, all charge impurities are assumed to reside at the same distance from graphene layer. Second, other parameters which are known to depend on the details of the substrate, such as the Fermi velocity, were fixed to accepted values. Third, the model considers Coulomb scattering alone. Although in the regime studied here this is the dominant scattering mechanism, other weaker mechanisms which come into play such as electron-phonon interactions or scattering from bulk and edge defects, have different energy and temperature dependence and may contribute to the deviation.